\documentclass[twocolumn,useAMS,usenatbib]{mn2e}

\usepackage{graphicx}% Include figure files
\usepackage{subfigure}% Include subfigure files
\usepackage{xcolor}% Change colors of text using \textcolor{red}{...}
\usepackage{mathtext,bm,bbm,amsmath,amsfonts,amssymb,indentfirst,syntonly,graphicx}
\usepackage{mathtools}
\usepackage{slashbox}
\usepackage[english]{babel}
\usepackage{calc}
\usepackage{tikz}
\usepackage[T1]{fontenc}
\usepackage{ae,aecompl}

\def\bc{\begin{center}}
\def\ec{\end{center}}
\def\be{\begin{eqnarray}}
\def\ee{\end{eqnarray}}

\title[Probing CDM with GW fringes detected by ET]{Probing compact dark matter with gravitational wave fringes detected by the Einstein Telescope}
\author[Liao et al.]
{Kai Liao$^{1}$\thanks{e-mail: liaokai@whut.edu.cn},
Shuxun Tian$^{2}$,
and Xuheng Ding$^{3}$\\
$^{1}$School of Science, Wuhan University of Technology, Wuhan 430070, China\\
$^{2}$School of Physics and Technology, Wuhan University, Wuhan 430072, China\\
$^{3}$Department of Physics and Astronomy, University of California, Los Angeles, CA, 90095-1547, USA}
\begin{document}

\date{Accepted xxxx; Received xxxx; in original form xxxx}

\pagerange{\pageref{firstpage}--\pageref{lastpage}} \pubyear{2018}

\maketitle

\label{firstpage}

\begin{abstract}
  Unlike the electromagnetic radiation from astrophysical objects, gravitational waves (GWs) from binary star mergers have much
  longer wavelengths and are coherent. For ground-based GW detectors, when the lens object between the source and the earth has
  mass $\sim 1-10^5M_\odot$, the diffraction effect should be considered since the chirping wavelengths are comparable to the scale of the barrier (its Schwarzschild radius).
  The waveform will thus be distorted as the fringes.
  In this work, we show that signals from the third-generation GW detectors like the Einstein Telescope (ET) would be a smoking gun for probing the nature of compact dark matter (CDM) or primordial black holes.
  Detection of the lensing effects becomes harder when the lens mass is smaller.
  ET is more sensitive than LIGO, the constraint is available for CDM mass $>5M_\odot$ while LIGO can only detect the mass $>100M_\odot$. For a null search of the fringes, one-year observation of ET can constrain the CDM
  density fraction to $\sim10^{-2}-10^{-5}$ in the mass range $M_{\rm{CDM}}=10M_\odot-100M_\odot$.
\end{abstract}


\section{Introduction}
Cosmological observations have indicated that a considerable part of the energy density in the current Universe consists of dark matter.
However, we still know little about its nature and composition. One may conjecture that dark matter (or part of it) exists in the form of compact objects. Theoretical models from particle physics and cosmology include the massive compact halo objects (MACHOs)~\citep{Wyrzykowski2011,Pooley2009,Mediavilla2009,Monroy-Rodriguez2014}, primordial black holes (PBHs)~\citep{Carr1974,Carr1975}, axion mini-clusters~\citep{Hardy2017}, compact mini halos~\citep{Ricotti2009} and so on. Hereafter we take all of them as the compact dark matter (CDM).
The mass of CDM could be as light as $10^{-7}M_\odot$ and as heavy as the first stars $\sim 10^3M_\odot$~\citep{Griest1991}.

Many efforts have been devoted to probing CDM with various approaches.
While the wide stellar binaries could be perturbed by large-mass CDM ($\geqslant100M_\odot$)~\citep{Quinn2009}, the microlensing observations of the stars can constrain
the CDM in our galaxy with low-mass ($\leqslant10M_\odot$)~\citep{Tisserand2007,Wyrzykowski2011,Udalski2015,Calchi Novati2013,Niikura2017}.
The cosmic microwave background (CMB) can also give constraints by lack of radiation as a result of accretion~\citep{Ali-Haimoud2017}.
Other methods include supernova lensing~\citep{Benton2007}, caustic crossing~\citep{Oguri2018}, ultra-faint dwarf galaxies~\citep{Brandt2016} and millilensing of quasars~\citep{Wilkinson2001}.
The observational constraints specifically focus on two parameters of CDM: the fraction $f_{\rm{CDM}}$ of dark matter and the mass $M_{\rm{CDM}}$.
Current constraints are generally quite weak for $f_{\rm{CDM}}\lesssim0.1$ and in some mass windows.

It is worth mentioning that the mass range $10 M_\odot\leqslant M_{\rm{CDM}}\leqslant100 M_\odot$ has being received lots of attention especially after LIGO/VIRGO published their first successful detection~\citep{Abbott2016}. It was pointed out that GW150914 might be a signature of PBH dark matter~\citep{Bird2016,Sasaki2016}. However, the existing astronomical constraints about the abundance of PBHs in this mass range are too weak~\citep{Ricotti2008,Oguri2018} and thus not sufficient to test this conjecture.
Independent and more powerful methods are quite needed.
New methods based on lensing of transient sources like the Gamma Ray Bursts (GRBs)~\citep{Ji2018}, the fast radio bursts (FRBs)~\citep{Munoz2016} were proposed to probe this mass window
since CDM as the lenses would alter the observed signals appearing as echoes.
Remarkably, lensing of FRBs is expected be very promising since the intrinsic duration of FRB is $\sim\mathrm{msec}$, comparable to the time delay caused
by the lens mass in this window.
However, it will not be easy to apply this method in real data due to the unknown properties of FRBs.
For example, the redshifts are quite uncertain for most of signals and it is degenerate between a split signal and the intrinsic structure.

Direct detection of GW by LIGO/VIRGO opened a new window for astronomy and cosmology~\citep{Abbott2016}. The lensing of GW has been attracting the eyes of the community, especially in recent years~\citep{Sereno2010,Ding2015,Fan2017,Collett2017,Liao2017,Wei2017,Li2018,Smith2018,Lai2018,Oguri2018a,Dai2018,Yang2019}.
GW is completely distinct from and complementary to the electromagnetic (EM) wave. The emitters of EM radiation are charged particles, so EM radiation is emitted
within small regions due to the overall charge neutrality. EM wave has short wavelengths and emits independently from each part of the source.
By contrast, GWs are emitted by the cumulative mass and the momentum of entire systems, so they have long wavelengths and are coherent.
Therefore, while lensing of light is described by geometric optics without considering the interference, lensing of GW should be described by wave optics in some cases~\citep{Nakamura1998,Takahashi2003,Takahashi2017,Liao2019}.
The lens is like the diffraction barrier which distorts the GW waveform as the fringes. Lensing of GWs observed by advanced LIGO (aLIGO) was proposed to probe the CDM as the lenses~\citep{Jung2019}.
While this idea is novel, the constraint power is quite weak for aLIGO, especially in the mass range $10 M_\odot\leqslant M_{\rm{CDM}}\leqslant100 M_\odot$~\citep{Jung2019}.
Motivated by this, in this work, we consider the fringes observed by the third-generation ground-based detectors, for example, the Einstein Telescope (ET).
We prove that ET could be the key to pin down some nature of dark matter.

The paper is organized as follows: In Section 2, we introduce the lensing of GW with wave optics description; In Section 3, we introduce the distributions
of binary star mergers and the ET; The methodology and results are presented in Section 4; Finally, we summarize and make discussions in Section 5. Throughout this paper, we use
the natural units of $c=G=1$ in all equations.

\section{Wave optics description}
For GW lensing, if the wavelength is much shorter than the lens mass scale, the GW travels along geodesic in the geometric optics limit as the light~\citep{Wang1996}.
In this case, the interference of two signals would occur~\citep{Hou2019}. On the contrary, if the wavelength is much longer than the lens, it belongs to the wave optics limit.
In the intermediate regime, diffraction effect should be considered, which is described by wave optics~\citep{Nakamura1998,Takahashi2003}. We consider
the gravity field of the CDM is weak, the metric is given by
\begin{equation}
g_{\mu \nu}=g^{(L)}_{\mu \nu}+h_{\mu \nu},
\end{equation}
where $g^{(L)}_{\mu \nu}$ is determined by the Newtonian potential ($U$) of the lens. $h_{\mu \nu}$ is the GW perturbation
which can be separated with a scalar field and its polarization:
\begin{equation}
h_{\mu \nu}=\phi e_{\mu \nu}.
\end{equation}
Since the gravity field is weak, the polarization is taken as constant. Thus we take GW as a scalar wave whose evolution is determined by
\begin{equation}
\partial_\mu(\sqrt{-g^{(L)}}g^{(L)\mu \nu}\partial_\nu\phi)=0.
\end{equation}
Using Newtonian potential, we rewrite it for given frequency $f$ as
\begin{equation}
(\Delta + 4\pi^2 f^2)\tilde{\phi}=16\pi^2f^2U\tilde{\phi}.
\end{equation}

To quantify the impacts of the lens on the waveform, we follow~\citep{Takahashi2003} to
define the dimensionless amplification factor as
\begin{equation}
F(f)=\tilde{\phi}^L(f)/\tilde{\phi}(f),\label{Ff}
\end{equation}
where $\tilde{\phi}^L$ and $\tilde{\phi}$ are the lensed and unlensed ($U=0$) amplitudes in frequency domain, respectively.
As diffraction, the observed signal is the superposition of all possible waves on the lens plane that have different time delays
corresponding to different phases. Therefore $F(f)$ is given by~\citep{Takahashi2003}
\begin{equation}
F(f)=\frac{D_sR^2_E(1+z_l)}{D_lD_{ls}}\frac{f}{i}\int d^2\mathbf{x}\ \exp\left[2\pi ift_d(\mathbf{x},\mathbf{y})\right]\label{amplitude},
\end{equation}
where $R_E$ is the Einstein radius, $\mathbf{x},\mathbf{y}$ are the impact position in the lens plane and source position in units of $R_E$.
$t_d$ is the relative travelling time for any path on the lens plane:
\begin{equation}
t_d(\mathbf{x},\mathbf{y})=\frac{D_sR^2_E(1+z_l)}{D_lD_{ls}}\left[\frac{1}{2}|\mathbf{x}-\mathbf{y}|^2-\psi(\mathbf{x})+\phi_m(\mathbf{y})\right],
\end{equation}
where $\psi(\mathbf{x})$ is dimensionless deflection potential and $\phi_m(\mathbf{y})$ is chosen such that the minimum arrival time is zero. The CDM can be regarded as a point mass, whose potential
$\psi(\mathbf{x})=\ln{x}$ and  $\phi_m(\mathbf{y}) = 0.5\;(x_m - y)^2 - \ln{x_m}$ with $ x_m = 0.5 \;(y + \sqrt{y^2 + 4})$.
In such case, Eq.\ref{amplitude} can be written as
\begin{align}
  F(f)&=\exp\left\{\frac{\pi w}{4}+\frac{iw}{2}\left[\ln(\frac{w}{2})-2\phi_m(y)\right]\right\}\nonumber\\
  &\quad\times\Gamma\left(1-\frac{iw}{2}\right)\,_1F_1\left(\frac{iw}{2},1,\frac{iw}{2}y^2\right),
\end{align}
where $_1F_1$ is the confluent hypergeometric function and the dimensionless parameter
$w=8\pi M_{\rm{CDM}}(1+z_l)f$, which
serves as a comparison between the barrier scale (Schwarzschild radius $R_s=2M_{\rm{CDM}}$) and the wavelength [$\lambda\simeq2\times10^3(100Hz/f)M_\odot$].

\section{Binary star mergers and the Einstein telescope}
\begin{figure}
 \includegraphics[width=8cm,angle=0]{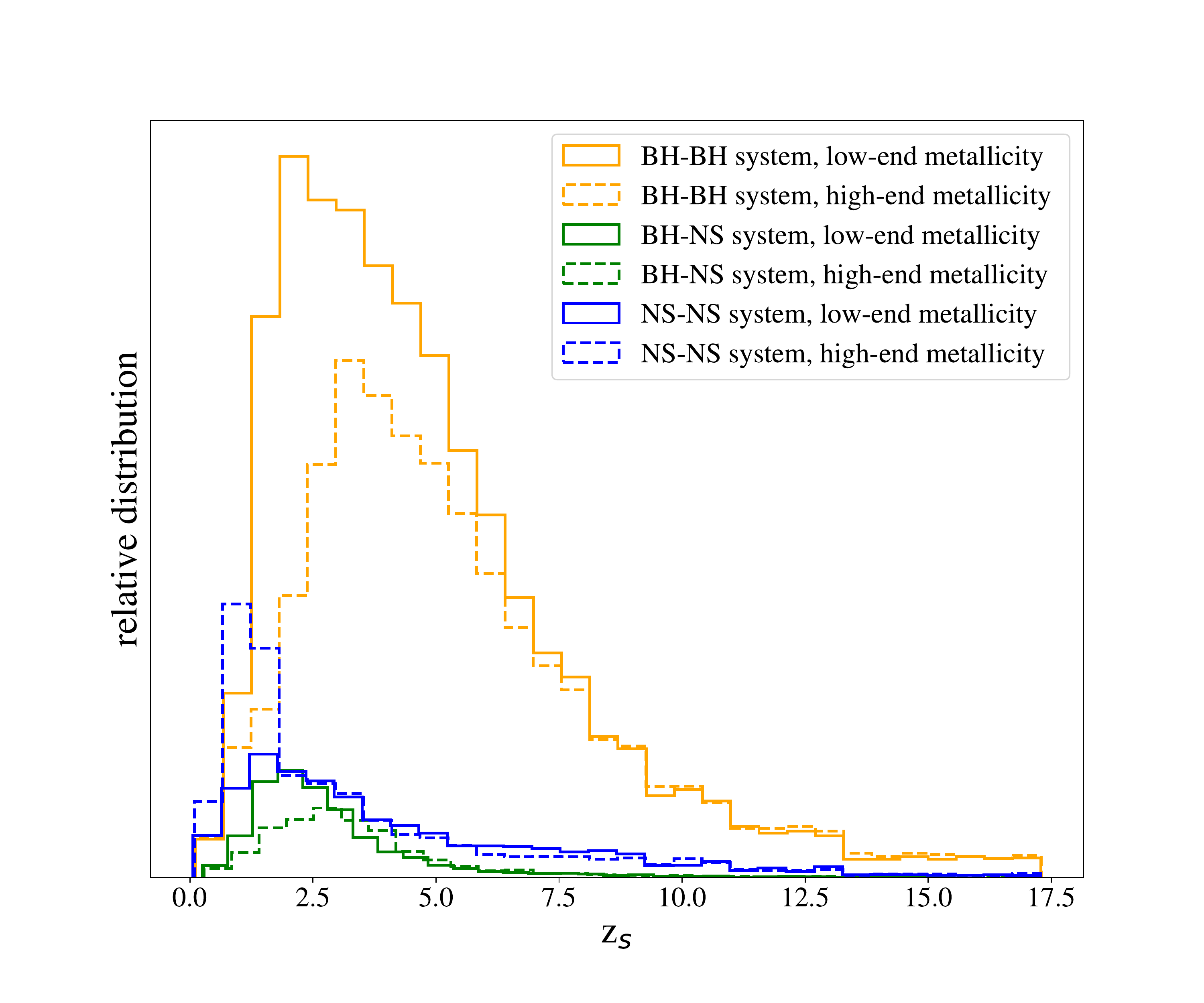}
   \caption{
   The relative distributions of arrived yearly merger rates for different classes of the double compact objects as a function of redshift for ``low-end" and ``high-end" metallicity evolutions,
   calculated by the \textit{StarTrack} evolutionary code.
  }\label{zs}
\end{figure}

Following our previous works~\citep{Ding2015,Liao2017}, we consider three classes of the dual compact objects including neutron star binaris (NS-NSs), black hole-neutron stars (BH-NSs) and black hole binaries (BH-BHs). We adopt the inspiral rates of these events reported by \citep{Dominik2013}, which contain the detailed population synthesis calculations predicted by the \textit{StarTrack} evolutionary code\footnote{https://www.syntheticuniverse.org/}. We consider the standard scenarios with two metallicity evolution conditions including ``low-end" and ``high-end" cases.
Figure~\ref{zs} illustrates the distributions of the arrived yearly merger events of the three classes as a function of redshift.
The simulated total event numbers of BH-BHs, BH-NSs and NS-NSs are 330641, 26865 and 47265 respectively for ``low-end", 244061, 19622 and 60346 respectively for ``high-end".
To be detected, they should further reach signal-noise-ratio (SNR) conditions.
For the BH-BH systems, the masses for each BH are considered in the following way. We follow the works by \citep{Kovetz2017, Fishbach2018, Abbott2019b} and assume the primary BH mass (i.e., $m_1$) follows a normalized power-law distribution~\citep{Ding2020}:
\begin{equation} \label{equ_powlaw}
P(m_1|\alpha, 80M_\odot, 5M_\odot) \propto m_1^{\alpha} \mathcal{H}(m_1-5M_\odot) \mathcal{H}(80M_\odot-m_1),
 \end{equation}
where $\mathcal{H}$ is the Heaviside step function. For the secondary BH, its mass (i.e., $m_2$) is sampled from a uniform distribution between $[5M_\odot, m_1]$.
For NS-NS and NS-BH systems, we take the average values $1.3M_\odot$ for NS mass and $13M_\odot$ for BH mass for simplicity.
Given that the NS-NS and NS-BH systems only cover a small fraction of the overall samples, our adoptions would not change the results much.

\begin{figure}
 \includegraphics[width=8cm,angle=0]{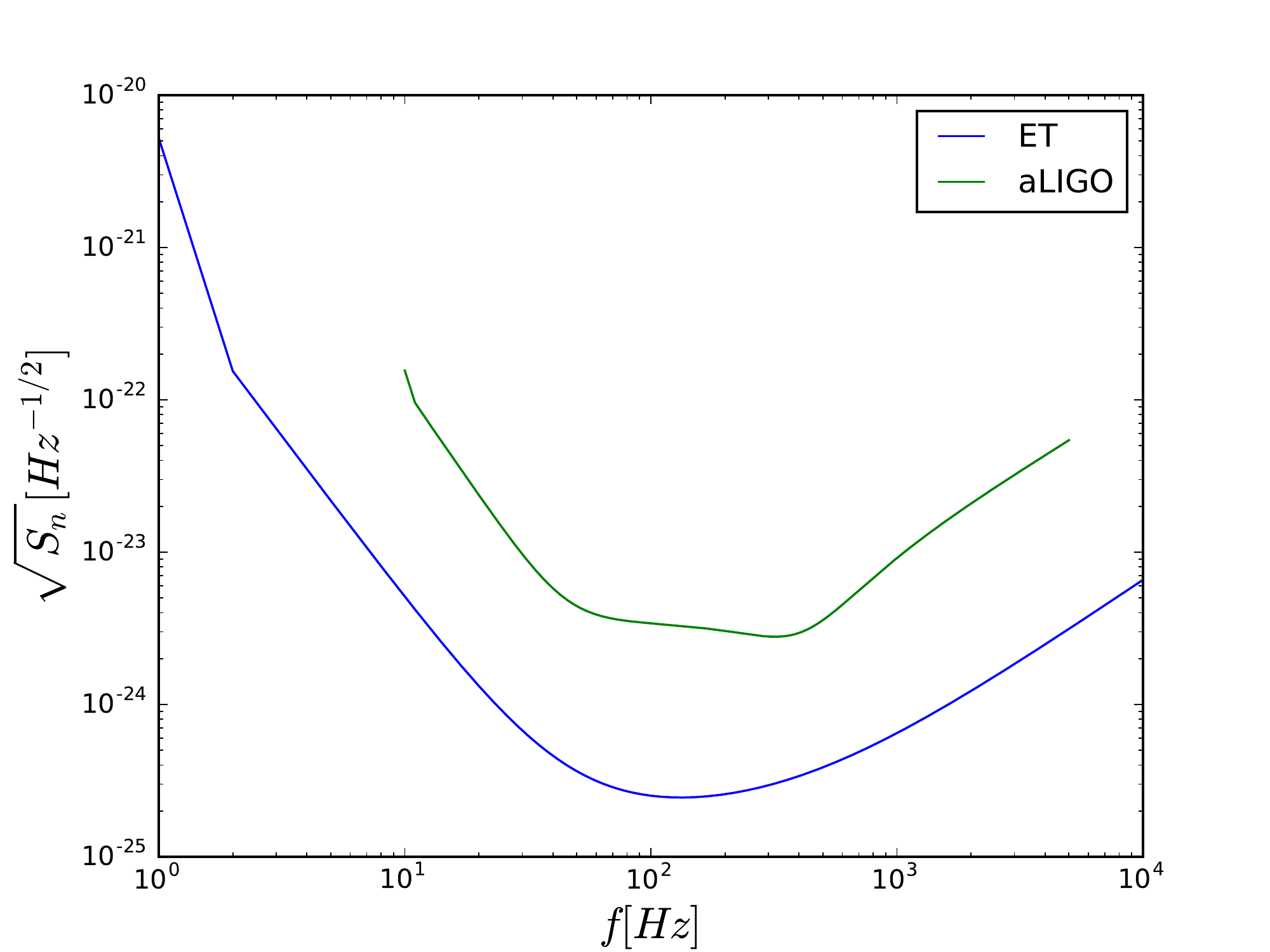}
   \caption{Sensitivity curves for aLIGO and ET.
  }\label{snr}
\end{figure}

The next-generation detectors like the Einstein Telescope will broaden the accessible volume of the Universe by three orders of magnitude promising tens to hundreds thousands of detections per year. ET consists of three Michelson interferometers with 10 km long arms, arranged to form
an equilateral triangles. Two sensitivity estimates have been put forward for ET. One is based
on a single interferometer covering the full frequency range 1 Hz - 10 kHz, referred as ET-B. The
other is the xylophone design (ET-D) in which one detector is composed of one cryogenic low-frequency interferometer and one room temperature high-frequency interferometer.
We adopt the ET-B in this work. The adoptions would not change the main conclusions.
The sensitivity curve of ET is approximately matched by \citep{Mishra2010}
\begin{align}
  &\sqrt{S_n(f)}=\sqrt{S_0}\left(2.39\times10^{-27}x^{-15.64}\right.\nonumber\\
  &\quad\left.+0.349x^{-2.145}+1.76x^{-0.12}+0.409x^{1.1}\right),
\end{align}
where $S_0=10^{-50}\,{\rm Hz}^{-1}$ and $x=f/(100\,{\rm Hz})$.
We also consider aLIGO for comparison~\citep{Mishra2010}. For intuition, we plot these two sensitivity curves in Fig. \ref{snr}.

\section{Methodology and results}

For the waveform from a binary star merger event detected by ground-based detectors,
we follow ~\citep{Jung2019,Cao2014}, ignoring higher-order post-Newtonian terms, spin effects, orbital eccentricity and nonquadrupole modes for simplicity.
The unlensed waveform in frequency domain is given by
\begin{equation}
  \tilde{h}(f)=\tilde{A}(f)\exp\{i[\Psi(f)+\phi_0]\},
\end{equation}
where
\begin{equation}
  \tilde{A}(f)=\sqrt{\frac{5}{24}}\frac{\mathcal{M}_z^{5/6}\mathcal{F}}{\pi^{2/3}d(z)}f^{-7/6},
\end{equation}
and
\begin{equation}
  \Psi(f)=2\pi ft_c+\frac{3}{128}\left(\pi \mathcal{M}_zf\right)^{-5/3},
\end{equation}
with redshifted chirp mass
\begin{equation}
  \mathcal{M}_z=(1+z)\frac{(m_1m_2)^{3/5}}{(m_1+m_2)^{1/5}}.
\end{equation}
$t_c$ is the coalescence time. $d$ is the luminosity distance. The angular orientation function $\mathcal{F}$ contains all angle dependence of the detector response to binary inspiral with \citep[see][for details]{Finn1996}
\begin{equation}
\mathcal{F}^2=(1+\cos^2\iota)^2F_+^2/4+\cos^2\iota F_\times^2.
\end{equation}
A random configuration gives the probability distribution \citep{Finn1996}
\begin{equation}
  P(\mathcal{F}) = \left\{ \begin{array}{ll}
    20\,\mathcal{F}(1-\mathcal{F})^3\  & {\rm if}\ 0<\mathcal{F}<1,\\
    0 & \textrm{otherwise}.
  \end{array} \right.
\end{equation}

The lensing optical depth for a given GW event at $z_s$ is the probability that the event falls into the perceptible region ($y<y_{\rm{max}}$)
of any CDM along the line of sight:
\begin{equation}
\tau(M_{\rm{CDM}}, n_L, \mathcal{S})=\int_0^{z_s}d\chi(z_l)(1+z_l)^2n_L\sigma(M_{\rm{CDM}},z_l,\mathcal{S}), \label{taudef}
\end{equation}
where $n_L$ is the CDM number density and the cross section is given by
\begin{equation}
\sigma(M_{\rm{CDM}},z_l,\mathcal{S})=4\pi M_{\rm{CDM}}\frac{D_lD_{ls}}{D_s}y_{\rm{max}}^2(M_{\rm{CDM}},z_l,\mathcal{S}),
\end{equation}
where $\mathcal{S}=\{M_z,d(z_s),\phi_0,\mathcal{F},S_n(f),f_0,f_1\}$ is determined by the GW source itself and the detector.
To determine $y_{\rm{max}}$, first of all, SNR must be large enough:

\begin{figure}
 \includegraphics[width=8cm,angle=0]{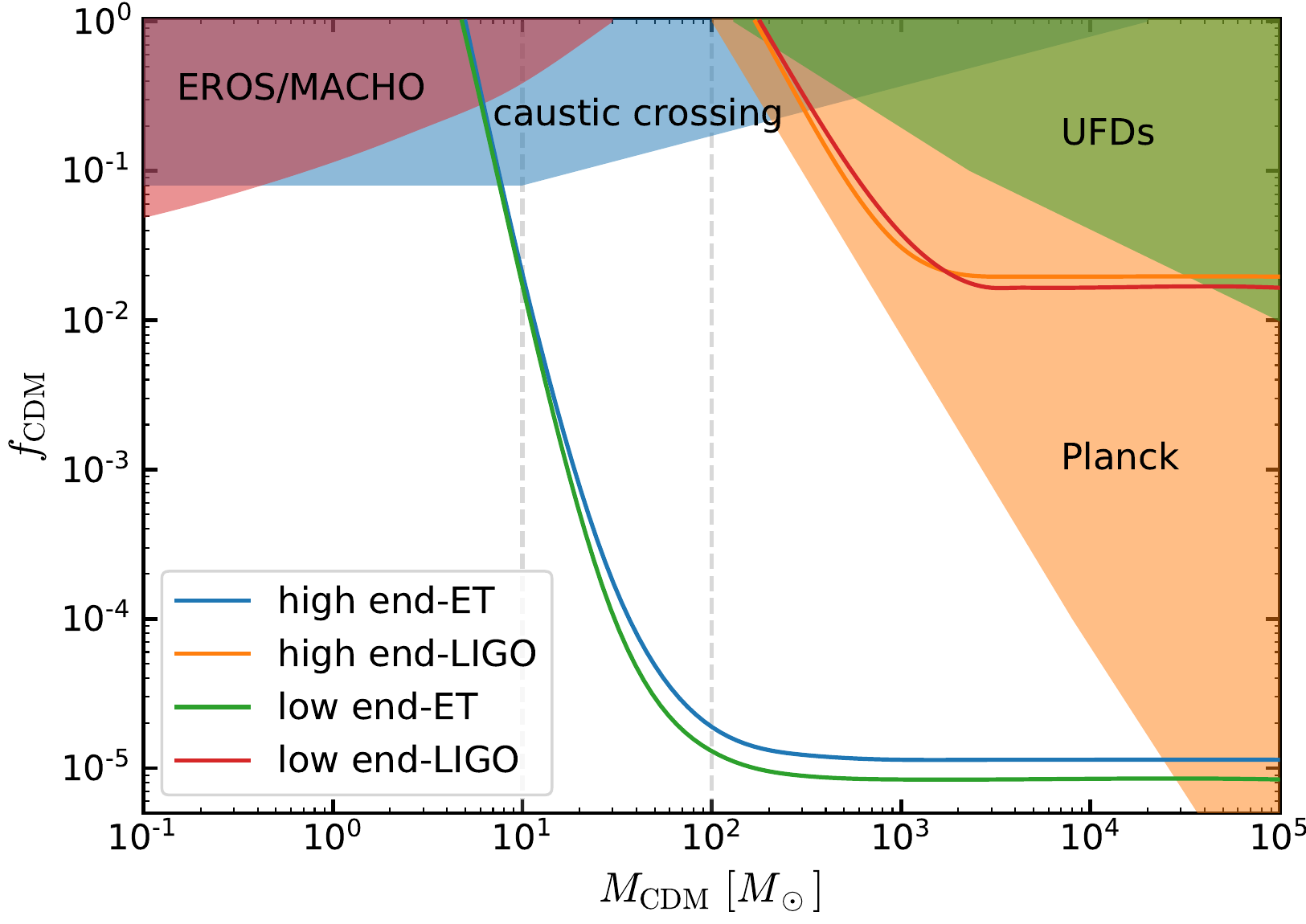}
   \caption{Constraints on the mass $M_{\rm{CDM}}$ and fraction $f_{\rm{CDM}}$ from one-year observation by aLIGO and ET. Shaded regions show the excluded parameter space from EROS/MACHO microlensing~\citep{Alcock2000,Tisserand2007}, caustic crossing \citep{Oguri2018}, Planck CMB observations \citep{Ali-Haimoud2017}, and ultrafaint dwarf galaxies (UFDs)~\citep{Brandt2016}, respectively.
   The constraint by ET is available for $M_{\rm{CDM}}>5 M_{\odot}$, i.e., for a null search of lensed signals, one can exclude the case that all dark matter consists of compact objects in this mass range.
  }\label{results}
\end{figure}

\begin{equation}
(\mathrm{SNR})^2 = 4\int_{f_0}^{f_1}\frac{|\tilde{h}(f)^L|^2}{S_n(f)}df,
\end{equation}
where the lower cutoff $f_0$ depends on the detectors \citep{Mishra2010},
we adopt $f_0=1{\rm Hz}$ and $f_0=10{\rm Hz}$ for ET and aLIGO, respectively.
The cutoff frequency $f_{\rm{cut}}=[3\sqrt{3}\pi (1+z)(m_1+m_2)]^{-1}$ \citep{Jung2019}.
We adopt $f_1=\rm{min}$$(f_{\rm{cut}},10{\rm kHz}$) for ET and $f_1=\rm{min}($$f_{\rm{cut}},5{\rm kHz}$) for aLIGO, respectively.
The lensed waveform $\tilde{h}(f)^L=F(f)\tilde{h}(f)$ according to the definition in Eq.\ref{Ff}.
To ensure we can observe the signal, we adopt $\mathrm{SNR}>8$ as our first criterion. Besides, the lensed signal should be distinct from unlensed one. One way to assess this is through parameter estimation~\citep{Cao2014}.
Equivalently, we can use the following criterion~\citep{Jung2019}:

\begin{equation}
(\mathrm{SNR_{test}})^2 = 4\int_{f_0}^{f_1}\frac{|\tilde{h}(f)^L-\tilde{h}(f)_{\rm{best-fit}}|^2}{S_n(f)}df.
\end{equation}
To ensure we can figure out the lensing signature, we adopt $\mathrm{SNR_{test}}>5$ as the second criterion.
We use the standard template ($\tilde{h}(f)$) to fit the lensed signal. The parameters allowed to vary are the amplitude and the phase.

If a signal is too weak ($\mathrm{SNR}<8$) or deviates too far from the lens center ($y>y_{\rm{max}}$) such that one can not distinguish the lensing effects ($\mathrm{SNR_{test}}<5$),
it can not be verified as a lensed signal. Therefore, for each event, its $y_{\rm{max}}$ is determined by simultaneously satisfying $\mathrm{SNR}>8$ and $\mathrm{SNR_{test}}>5$, as a function of
$M_{\rm{CDM}}$, $z_l$ and $\mathcal{S}$. The status of $y_{\rm{max}}$ in this work is similar to the Einstein radius with which one can calculate the optical depth in the traditional microlensing.

Using Hubble parameter, we can rewrite Eq.\ref{taudef} as:
\begin{equation}
\begin{aligned}
\tau(M_{\rm{CDM}},f_{\rm{CDM}},\mathcal{S})=\frac{3}{2}f_{\rm{CDM}}\Omega_c\int_0^{z_s}dz_l\frac{H_0^2}{H(z_l)}\frac{D_lD_{ls}}{D_s}    \\
\times(1+z_l)^2y_{\rm{max}}^2(M_{\rm{CDM}},z_l,\mathcal{S}).  \label{tau}
\end{aligned}
\end{equation}
We adopt the flat $\Lambda$CDM model with total dark matter density $\Omega_c=0.24$, baryonic matter density $\Omega_{\rm{b}}=0.06$ and Hubble constant $H_0=70\rm{km\ s^{-1}Mpc^{-1}}$ for the simulation.
Following all the works in the literature, we assume a fraction of dark matter is in the form of compact dark matter which has the same mass $M_{\rm{CDM}}$, the density is $\Omega_{\rm{CDM}}$, then $f_{\rm{CDM}}=\Omega_{\rm{CDM}}/\Omega_c$.

According to the definition of $\tau$ for each event, i.e., the probability of verifying a lensed GW, the anticipated number of lensed GWs is the sum of
all detected GW events by ET in one year:
\begin{equation}
N_{\rm{lensed}}(M_{\rm{CDM}},f_{\rm{CDM}})=\sum_{i=1}^{N_{\rm{total}}}\tau_i.
\end{equation}
For a null search of lensed GW signals, the region in the ($M_{\rm{CDM}},f_{\rm{CDM}}$) space that predicts at least one detectable lensed event ($N_{\rm{lensed}}\geq1$) should be ruled out.
Our pipeline calculates $N_{\rm{lensed}}$ for each point in Fig.\ref{results} and finds the critical curves corresponding to $N_{\rm{lensed}}=1$.

The constraint results are shown in Fig.\ref{results}.
If no lensed signal is found by ET, the parameter space above the curves
will be excluded. As one can see, ET is able to constrain the mass as small as $\sim5M_\odot$, consistent with ~\citep{Christian2018}. For mass range $M_{\rm{CDM}}=10M_\odot-100M_\odot$,
$f_{\rm{CDM}}$ will be $10^{-2}-10^{-5}$, and for mass $>100M_\odot$, it gradually gets close to $\sim10^{-5}$, which is the best constraint among the current methods.
For comparison, we also consider aLIGO. The constraints are much weaker, and it is unavailable for mass $<100M_\odot$.
Our results are consistent with  ~\citep{Jung2019}, though they claimed $M_{\rm{CDM}}>20M_\odot$ is detectable
at the level of $f_{\rm{CDM}}\sim0.1$ for $\rm{SNR}_{\rm{test}}>3$ and for the optimistic black hole model (see Fig.5 therein).
It would be interesting to compare our results with theirs since we use different binary star models.
Despite of the small differences, all the results should manifest that aLIGO is quite weak for this mass window.
Therefore, for the purpose of detecting CDM in this range, ET will be more promising.
We also compare our results with current constraints.

\section{Conclusion and discussions}
Whether dark matter exists in the form of compact objects is one of the most interesting problems for physicists and cosmologists.
Compact dark matter could play the role of the lens, leaving the fringes in the detected gravitational waves.
We have proved that its mass and fraction can be well understood especially in the mass range $M_{\rm{CDM}}=10M_\odot-100M_\odot$ by the Einstein Telescope.
The density fraction can be constrained at the level of $\sim10^{-2}-10^{-5}$ in such range for a null search of lensed signals.
It should be easy to add the lensing effects to the standard waveform templates for searching GWs.
Constraints in this mass window would supplement the knowledge on the nature and origin of dark matter.

In this work, note that a constant mass function was assumed as all the analysis did in the literature. In fact, CDM may have its own mass function, for example, a power-law form.
Further works combining all the methods need to be done to distinguish the mass functions.

Our analysis is from a statistical point of view. We only take the information that a GW signal is lensed or not. Actually, once a lensed signal is confirmed,
further information of the lens mass can be extracted based on the parameter estimation.
The redshifted lens mass can be well constrained~\citep{Cao2014}, thus the mass of CDM can be pinned down on a certain order of magnitude with an uncertainty coming from the lens redshift ($0<z_L<z_S$),
further reducing the allowed parameter space of CDM.

In addition to ET, another proposed US-based third-generation GW detector is Cosmic Explorer (CE)
which will keep L-shaped configuration with 40 km arm lengths. A network of two or more third-generation detectors would further enhance the constraints.

\section*{Acknowledgments}
This work was supported by the National Natural Science Foundation of China (NSFC) No. 11973034.

\label{lastpage}

\end{document}